# Polarization in Decentralized Online Social Networks

Lucio La Cava
DIMES Dept., University of Calabria
Rende, Italy
lucio.lacava@dimes.unical.it

Domenico Mandaglio
DIMES Dept., University of Calabria
Rende, Italy
d.mandaglio@dimes.unical.it

Andrea Tagarelli
DIMES Dept., University of Calabria
Rende, Italy
andrea.tagarelli@unical.it

## ABSTRACT

Centralized social media platforms are currently experiencing a shift in user engagement, drawing attention to alternative paradigms like Decentralized Online Social Networks (DOSNs). The rising popularity of DOSNs finds its root in the accessibility of open-source software, enabling anyone to create a new instance (i.e., server) and participate in a decentralized network known as Fediverse. Despite this growing momentum, there has been a lack of studies addressing the effect of positive and negative interactions among instances within DOSNs. This work aims to fill this gap by presenting a preliminary examination of instances' polarization in DOSNs, focusing on Mastodon — the most widely recognized decentralized social media platform, boasting over 10M users and nearly 20K instances to date. Our results suggest that polarization in the Fediverse emerges in unique ways, influenced by the desire to foster a federated environment between instances, also facilitating the isolation of instances that may pose potential risks to the Fediverse.

## KEYWORDS
Fediverse, Polarization, Mastodon, Signed Network

## 1 INTRODUCTION

Centralized social media platforms are nowadays facing a transformative shift, pushing users to seek new social interaction cues. In this regard, an increasing number of users are turning their attention toward alternative paradigms, with Decentralized Online Social Networks (DOSNs) [8, 11] emerging as prominent contenders [13, 15, 19].

The allure of DOSNs relies upon the availability of open-source software, allowing individuals to create *instances* (i.e., servers) and seamlessly interact with other instances, similarly to email services, in a decentralized network known as the *Fediverse*.

Among the plethora of decentralized services developed in recent years, Mastodon stems as the most widely recognized and studied [3, 23, 24, 26, 27] decentralized social media, boasting a user base exceeding 10M and nearly 20K instances to date, thus serving as the best representative for delving into the Fediverse.

The steady growth of Mastodon users has fostered the proliferation of social interactions that are articulated on two relational fronts: interactions among users spread across different instances, and interactions among instances deriving from the former. Existing studies mainly focused on characterizing the typical relational front, i.e., those among users. In this regard, the followers-followee distribution on Mastodon is found to be more balanced than on Twitter [27], with a lower presence of social bots and unusual disassortative traits. The impact of decentralization of user relations has been also investigated, unveiling the emergence of strategic user roles [21] and characterizing the information consumption-production dichotomy in such a novel social context [22]. Conversely, the novel relational front among instances has been little investigated, with the unveiling and characterization of the structural footprint of the network of Mastodon instances [20].

However, despite such a body of studies has provided considerable insights into the decentralized social ecosystem, the semantics associated with relationships — being them among users or instances — was overlooked. Indeed, users establish *positive* relationships such as friendships, agreements, and supports, as well as *negative* relationships such as foes, disagreements, and distrust. The existence of such mixed interactions has led to an ever-growing *polarization* phenomenon, i.e., a division of the set of users into groups with opposite views on controversial topics (e.g., politics, religion, sport).

In the past few years, we have witnessed a plethora of studies about user polarization on social media [5, 6, 9, 10, 14, 17]. Interestingly, in DOSNs, we can also witness instances establishing positive relationships — which involve cooperation among servers, shared topics, common language, etc. — and negative relationships, such as those involving banning, blocking, moderation. Consequently, the phenomenon of polarization on a DOSN like Mastodon extends beyond individual users, manifesting at the instance level as well.

To date, however, the exploration of polarization phenomena, both at the user and instance level within DOSNs remains an underresearched area, lacking comprehensive analysis or empirical investigation. Understanding polarization in Mastodon is crucial as it reveals how communities form and interact within decentralized networks, guiding interventions to foster healthier online environments and mitigate negative effects on user experience.

**Contributions.** This work aims to address the aforementioned gap in the literature by conducting the first investigation of the polarization phenomena at the instance level within Mastodon. The choice to focus on instances is all but arbitrary, as gaining a macro-perspective of the polarization in Fediverse serves as a key precursor for the user level, yet still bears significant implications for the underlying user base.

To characterize polarization in DOSNs, we aim to answer the following research questions:

**(RQ1)** How many polarized groups can be found in Mastodon?

**(RQ2)** What is the polarization structure in Mastodon, that is, how are polarized groups linked internally and to each other?

**(RQ3)** What are the main characteristics of the instances within the detected polarized groups?

To answer the above RQs, our approach is built on (*i*) modeling the interactions among Mastodon instances through *signed graphs* where edges are labeled positively or negatively, indicating whether the connection they represent is friendly or antagonistic, respectively, and (ii) adopting existing methodologies for detecting polarization phenomena on signed graphs.

Our methodology builds on the body of studies formulating the problem of detecting polarization as partitioning the node set in a signed graph into $k$ non-overlapping sets, generally referred to as groups or poles, of the input signed graph [4, 7, 18]. In particular, following previous methods [2, 25], we recognize that the identified groups do not necessarily need to cover the entire node set, thus complying the most with real-world situations allowing neutrality.

## 2 DATA

Due to Mastodon's decentralized and scattered footprint, identifying the plethora of currently existing instances is a complex task. However, for obtaining the latest information about the Mastodon instance landscape, the instances.social[1] website is widely utilized as a de-facto tracker. To start building our data, we leveraged this platform to compile a seed set of instances to investigate, corresponding to those active at the time of writing this work. We thus explored such instances to define a seed set of around 270K Mastodon users.

Subsequently, we conducted a *breadth-first* search over them, identifying both incoming and outgoing links (via the /api/v1/acco unts/:id/followers and /api/v1/accounts/:id/following endpoints). As part of this process, already having been proven meaningful in [20], we systematically expanded the user exploration by incrementally discovering new users through the link detection mechanism. We let this crawling process go for the past nine months and eventually came up with more than 2M users and 116M unique links among them.

To the aim of our work, we used the following relations to infer the **positive links among instances**, deriving from social interactions between their users, as outlined in the following section.

Furthermore, to define the **negative links among instances**, we leveraged the /api/v1/instance/domain_blocks endpoint, which enables the detection of the set of moderated instances through a list of *DomainBlocks* objects,[2] i.e., JSON files containing blocked instances and associated metadata (e.g., the severity and the motivation of the block). We crawled blocked domains from our seed set of instances between July and November, obtaining more than 135K raw enforced blocks among instances.

We emphasize that according to the concept of federation underlying DOSNs, Mastodon users, and hence their instances, can engage seamlessly with other services in the Fediverse that adopt the same protocol (i.e., *ActivityPub*). Consequently, our crawled links also encompass interactions from/to other services within the Fediverse (e.g., Pleroma). However, these might provide valuable insights into our first exploration of polarization within DOSNs, and to this aim, we have chosen not to filter them out.

## 3 METHODOLOGY

**Network Modeling.** Let us denote with $\mathcal{U}$ the set of users and $\mathcal{I}$ the set of instances extracted from our data. We can infer a graph modeling positive relations among instances in $\mathcal{I}$ as a directed-weighted network $\mathcal{G}^+ = \langle V^+, E^+, w \rangle$, where $V^+ \subseteq \mathcal{I}$ is the set of nodes, $E^+$ is the set of *positive* edges such that $(i, j) \in E^+$ means that there exists at least one user in instance $i$ that follows a user in instance $j$, and $w : E^+ \mapsto \mathbb{R}$ is an edge weighing function such that, for any $(i, j) \in E^+$, $w(i, j)$ indicates the number of users from instance $i$ following users in instance $j$. Our network $\mathcal{G}^+$ contains 37,529 nodes and 1,335,490 edges. Similarly, we infer a graph modeling negative relations among instances in $\mathcal{I}$ as a directed network $\mathcal{G}^- = \langle V^-, E^- \rangle$, where $V^- \subseteq \mathcal{I}$ is the set of nodes, $E^-$ is the set of *negative* edges such that $(i, j) \in E^-$ means that instance $i$ banned instance $j$ during our crawling period. Our network $\mathcal{G}^-$ contains 11,401 nodes and 105,465 edges.

Before shaping our signed graph model for identifying polarized groups within Mastodon and the broadest Fediverse, we conducted a *network simplification task* aimed at pruning noisy or statistically irrelevant edges from our $\mathcal{G}^+$. Indeed, $E^+$ might also contain interactions among instances that are driven by spurious interactions by a pair of users within such instances. This is all but desirable, as it would impact our identification step. Therefore, we resorted to the theoretically well-rooted *Disparity Filter* [28] method, which exploits a *generative null model* built on node strengths to prune networks from statistically irrelevant edges based on *p*-value computation and specific significance thresholds. Specifically, the null hypothesis underlying the Disparity Filter states that the strength of a node is redistributed uniformly at random over its incident edges, thus evaluating the strength and degree of each node locally. We utilized the official Python implementation of Disparity Filter,[3] to filter $E^+$ removing irrelevant edges based on a significance threshold of $\alpha = 0.05$. This eventually results in a set of 117,422 edges in $E^+$. By contrast, we point out that the simplification step is not required for $\mathcal{G}^-$, as blocks among instances are explicitly declared by instances' administrators and not due to randomness.

Finally, we created our **signed instance-network** $\mathcal{G} = \langle V, E, s \rangle$, where $V \subseteq \mathcal{I}$, $E = E^+ \cup E^-$, and $s : E \mapsto \{+1, -1\}$ is a function that assigns each edge $(i, j) \in E$ to the corresponding sign, such that $w(i, j) = +1$ if $(i, j) \in E^+$, $-1$ otherwise. Our network $\mathcal{G}$ contains contains 19,738 nodes (i.e., instances) and 222,887 (positive/negative) relations among them. We emphasize that, during this merging step, we also accounted for the removal of ambiguous links, i.e., those existing with both signs due to the existence of a subsequent ban enforced toward instances already spotted as "positive" during our breadth-first search.

**Identification of Polarized Groups.** Let $E^+(P_i, P_j)$ (resp. $E^-(P_i, P_j)$) be the set of positive (resp. negative) edges between two subsets $P_i, P_j \subseteq V$. We define $E^+(P_i)$ (resp. $E^-(P_i)$) to be $E^+(P_i, P_i)$ (resp. $E^-(P_i, P_i)$) as the set of positive (resp. negative) intra-group edges, for any $P_i \subseteq V$. For a set of groups $\mathcal{P} = \{P_1, \cdots, P_k\}$ as a candidate solution, [25] quantifies the goodness of the solution by maximizing a function of the intra/inter-group connectivity as follows:

$$f(P_1, \cdots, P_k) = \sum_{P_i \in \mathcal{P}} \left( |E^+(P_i)| - |E^-(P_i)| \right)$$
$$+ \frac{1}{k-1} \sum_{P_i, P_j \in \mathcal{P}} \left( |E^-(P_i, P_j)| - |E^+(P_i, P_j)| \right).$$

---

It should be noted that the approach proposed in [25] is, to the best of our knowledge, the only existing method discovering an arbitrary number $k$ of polarized groups that admits *neutral nodes*, i.e., nodes that are not assigned to any of the polarized groups. The latter is an essential requirement to our problem under study, which significantly differs from conventional community detection and graph clustering problems. Please also note that comparing different algorithms for detecting polarized groups on Mastodon is beyond the scope of this work.

The identification of polarized groups is based on the following optimization problem, which is a special instance of the correlation clustering problem [1].

**PROBLEM 1 (K-CONFLICTING GROUPS [25]).** *Given a signed graph $G$ and an integer $k$, find $k$ mutually-disjoint node sets $\mathcal{P} = \{P_1^*, \cdots, P_k^*\}$ such that:*

$$P_1^*, \cdots, P_k^* = \underset{P_1, \cdots, P_k \subseteq V}{\arg\max} \frac{f(P_1, \cdots, P_k)}{|\cup_{i=1}^k P_i|} \qquad (1)$$

Under the formulation of Problem 1, the quality of the solution depends only on the polarized groups, and not on the neutral group $P_N = V \setminus \cup_{i=1}^k P_i$, which fulfills the above stated requirement. The proposed approach in [25] relies on interpreting the problem objective in terms of the Laplacian of a complete graph, characterizing the spectral properties of this matrix, and identifying each conflicting group as the solution to a maximum *Discrete Rayleigh Quotient* (DRQ) problem [25]. More specifically, the objective function in Eq. 1 is found in [25] as a convex combination of $k-1$ DRQ problems, whose solution to the $i$-th DRQ problem characterizes the group $P_i$ that conflicts the most with the remaining groups $P_j$, for $j > i$, yet to be investigated. The intensity of such conflict, characterizing group $P_i$, is referred to as the *DRQ value*. This value, having similar rationale to Eq. 1, should be maximized. Based on this observation, [25] proposes an iterative algorithm, dubbed SCG (Spectral Conflicting Groups), executing $k-1$ iterations that involve solving a DRQ problem.

## 4 RESULTS

**Determining the number of polarized groups $k$.** We adopted the heuristic proposed in [25], which is analogous to the "elbow" heuristic for $k$-means-like clustering methods. This involves the following steps: ($i$) executing the SCG method by different $k$ values, ($ii$) producing a plot of the DRQ values from each run, arranged in ascending order, plotting the $i$-th largest DRQ value at the $i$-th position, and ($iii$) determining $k$ to be one of the discernible "knees" on the resulting curve. We employed this heuristic by varying $k$ from 2 to 10 by increments of 1. For each $k$ value, we executed SCG 10 times and averaged the results, which are shown in Figure 1.

Our analysis indicates the presence of 4 conflicting groups in our signed instance-network, as the most distinct inflection point emerges during the 3rd iteration for multiple $k$ values. However, running SCG with $k = 4$ resulted in one empty polarized group, leaving us with 3 actual polarized groups that, alongside the neutral group $P_N$, will be the focus of our subsequent analysis.

**Characterizing main traits of the polarized groups.** We started gaining insights into the obtained polarized groups by means of the statistics shown in Figure 2 (top-left). Groups are all but evenly distributed, as the neutral group (or pole) $P_N$ encompasses more than 97% of the instances in our signed instance-network. This predominance of neutral instances perfectly aligns with the idea of Fediverse, supporting the co-existence of independent yet cooperating instances. Nonetheless, as previously mentioned, the Fediverse embraces multiple services, and we found evidence of such a feature in our groups. Indeed, while we spotted some almost *Mastodon-pure* groups (i.e., the non-neutral $P_1$ and $P_3$), the remainder (i.e., the neutral $P_N$ and the non-neutral $P_2$) is characterized by a fraction of Mastodon instances that does not exceed half of the group.

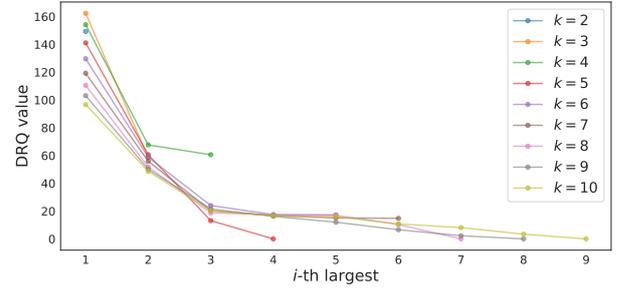

**Figure 1: DRQ plot from SCG runs by varying $k$.**

We also investigated the extent of "negative" edges (i.e., moderation actions) toward our spotted polarized groups, discovering that $P_N$ and $P_2$ stem as the groups collecting more bans within the Fediverse. However, while delving into the average number of bans within each group, we found that $P_1$ and $P_3$ receive a very small amount of bans, whereas the neutral group $P_N$ almost doubles the bans of the former groups, despite remaining on admissible values. Remarkably, the group $P_2$ emerges with a striking number of bans, which is almost $16x$, resp. $26x$, the average amount of received bans by $P_N$, resp. $P_1$ and $P_3$, hence stemming as the *ban-sink* pole of the Fediverse. Further confirmation of this trait emerges from the percentage of instances having received at least a ban within each pole: in fact, while the neutral and Mastodon-pure poles present a percentage of instances with at least one ban not exceeding half of the entire group (and in the best case, a quarter), the ban-sink pole $P_2$ stands out with a remarkable trait, with all encompassed instances having received at least a ban.

**Exploring relations between the polarized groups.** Once we identified the poles in the Fediverse, we focused on shaping the positive flows (i.e., interactions) and negative flows (i.e., bans) between their instances. As illustrated in Figure 2 (center), we observe that most interactions involve or are directed to the neutral group $P_N$, further corroborating the idea of neutrality that, in principle, characterizes the Fediverse. The remainder is split toward the Mastodon-pure groups, while the ban-sink one only receives interactions from itself. This isolation from interactions that characterize $P_2$ hints at a form of segregation toward its negative shade.

As concerns the negative flows, shown in Figure 2 (right), we unveiled an intriguing bipartite mode as bans appear to be split between $P_N$ and $P_2$. While the high number of incoming bans in the neutral group is affected by its large size, this does not hold for the ban-sink one, which appears once again to be segregated within the

|  | $(P_N)$ | $P_1$ | $P_2$ | $P_3$ |
|---|---|---|---|---|
| # Instances | 19,241 | 189 | 122 | 186 |
| % Mastodon | 43.6 | 92.6 | 36.1 | 91.4 |
| # Incoming bans | 79,690 | 728 | 24,651 | 396 |
| Avg. # bans | 12.94 | 7.35 | 202.06 | 7.62 |
| % Instances ≥ 1 ban | 32.0 | 52.4 | 100 | 28.0 |

|  | Most interacted | Most banned |
|---|---|---|
| $(P_N)$ | mstdn.jp | geofront.rocks |
| $P_1$ | mastodon.social | botsin.space |
| $P_2$ | pawoo.net | poa.st |
| $P_3$ | det.social | aethy.com |

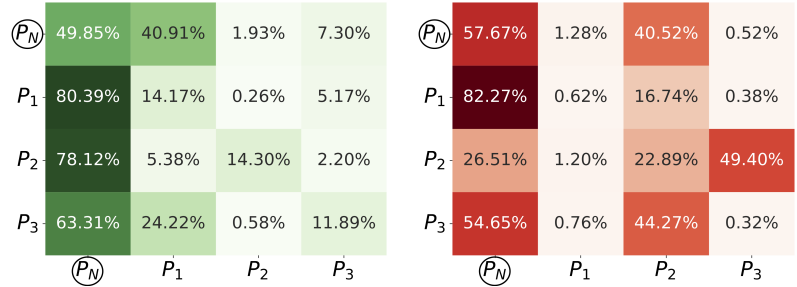

**Figure 2: (Top-left) Main characteristics of the polarized groups. (Bottom-left) Most interacted and banned instances for each group. (Center/Right) Positive/Negative flows of interaction among groups. Normalized percentages are to be read row-wise.**

|  | Volume | Avg | Top-active Instance | % Volume |
|---|---|---|---|---|
| $(P_N)$ | $2.37 \times 10^7$ | 3,654 | mstdn.jp | 7.65% |
| $P_1$ | $2.36 \times 10^7$ | 139,511 | mastodon.social | 32.84% |
| $P_2$ | $1.74 \times 10^6$ | 158,065 | pawoo.net | 54.55% |
| $P_3$ | $2.38 \times 10^6$ | 14,356 | mstdn.ca | 15.19% |

**Table 1: Pole activity (i.e., created posts) in the last 12 weeks, and corresponding most active instance.**

Fediverse. As a side yet relevant remark, while the Mastodon-pure group $P_1$ almost does not receive bans, $P_3$ receives almost half of the bans enforced from $P_2$, warranting further exploration.

**Unveiling main representative instances in polarized groups.** To further shed light on the group division that emerged above, we delved into each pole to unveil the most representative instances. Figure 2 (bottom-left) summarizes the most interacted, resp. banned, instances obtained through the in-strength of the positive edges, resp. in-degree of the negative ones. Below we discuss the most interesting instances we spotted. As concerns the neutral $P_N$, mstdn.jp emerges as the instance receiving the most positive interactions; this is not surprising as it turns out to be one of the earliest instances embracing the Mastodon rise, as well as the second largest Japanese instance in the Fediverse. The Mastodon-pure group $P_1$ sees mastodon.social as the most relevant positive instance, i.e., the official and first instance of the Mastodon project, whereas botsin.space arises as the most banned. For the latter, the reason is self-explanatory, as it hosts and runs Mastodon bots. Valuable insights emerge from the ban-sink group $P_2$. Indeed, the most interacted instance turns out to be pawoo.net, i.e., the second largest Mastodon instance in terms of users, recently under the spotlight due to the hosting of controversial content. The most banned instance in $P_2$ is poa.st, a non-Mastodon instance that advertises itself as the "Fediverse for shitposters",[4] further corroborating the negative connotation of that group of instances. Finally, another noteworthy trait involves the most banned instance in $P_3$, namely aethy.com, a server hosting potentially NSFW content.

**Activity in polarized groups.** We also characterized polarized groups in terms of activity via the `/api/v1/instance/activity`

endpoint of the Mastodon API, by collecting for each pole, the number of statuses created in the last 12 weeks (up to the time of writing this paper). As reported in Table 1, despite the large volume produced from all groups, two of them stand out, namely the neutral $P_N$ and the Mastodon-pure $P_1$. Nonetheless, the disparity in the average number of posts per instance within each group suggests the presence of a large tail of small instances within $P_N$, and identifies $P_1$ as the "beating core" of the Fediverse, as further confirmed by the presence of mastodon.social as the most prolific instance, covering one-third of the entire statuses created in $P_1$. Notably, the high average number of posts produced in the group $P_2$, coupled with the dominant role of pawoo.net (i.e., the instance producing more than half of the content) demand further investigation for unveiling the roots of such anomaly.

To this aim, we analyzed the most frequently used keywords for banning instances within our identified groups. As illustrated in Figure 3, and matching our previous insights from Figure 2 (left) witnessing the groups $P_1$ and $P_3$ as the least banned ones, we do not spot particularly evident motivations in their bannings. Conversely, valuable observations emerge from the remaining poles. Indeed, the ban-sink $P_2$ appears to be frequently banned for "speech" (hinting at hate-speech), "racism", and "harassment", providing additional shreds of evidence to our hypothesis of negativity. Finally, the most used ban keywords toward the neutral pole fuel the idea of federation, as "fedi" and "federate" emerge, suggesting the willingness to moderate instances that federate with unwelcome ones. The case in point is the presence of "facebook/meta" among the top-5 keywords, as a result of the emergence of *Threads*, the new social platform by Meta, as a potential "threat" to the Fediverse, following the desire to adopt the ActivityPub protocol.[5]

## 5 DISCUSSION

We conceived a signed network model aimed at detecting polarized groups in the Fediverse through the lens of Mastodon. Based on this model, we carried out the first explorative analysis of polarization phenomena among instances. Below we report the main findings that emerged through answering our research questions:

---

[4]https://globalextremism.org/post/poast/

[5]https://about.fb.com/news/2023/07/introducing-threads-new-app-text-sharing/

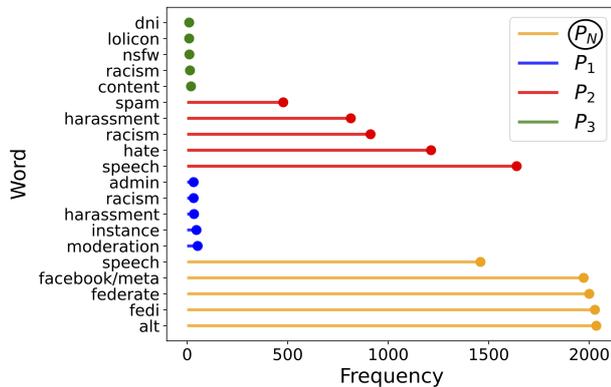

**Figure 3: Top-5 most frequent ban reasons for each pole.**

- **(RQ1)** The Mastodon-centric Fediverse instance network encompasses four non-overlapping groups of instances identified as poles.
- **(RQ2)** We unveil a unique polarization structure, featuring a predominant neutral group encompassing also instances from other services in the Fediverse, matching with the idea federation. The remaining instances are divided into two *Mastodon-pure* groups and a *ban-sink* group. Notably, positive interactions are predominantly observed between the neutral and pure groups, while the ban-sink group consistently receives negative links, as a protective measure against potential harm to the Fediverse.
- **(RQ3)** The Mastodon-pure group containing the most representative instance of the Fediverse (`mastodon.social`) and the neutral group exceed the other two groups in terms of created content volume. Besides, the *ban-sink* group exhibits anomalous trends in content production, characterized by particularly strong moderation due to the presence of harmful and inappropriate content.

Future work might delve into user-level polarization in DOSNs, comparing different algorithms for identifying polarized groups while also considering fairness aspects [12] and other ethical dimensions [16] throughout the detection process of polarized groups.